\documentclass[amsmath,amssymb,superscriptaddress,reprint]{revtex4-1}
\usepackage{graphicx}
\usepackage[dvipdfmx]{color}
\usepackage{dcolumn}
\usepackage{bm}
\usepackage{txfonts}
\usepackage{multirow}
\usepackage[utf8]{inputenc}
\usepackage[T1]{fontenc}
\usepackage{mathptmx}
\usepackage{etoolbox}
\usepackage[hypertex,colorlinks=true,linkcolor=black,citecolor=blue,filecolor=blue,urlcolor=blue,setpagesize=false,nesting=true]{hyperref}

\makeatletter
\def\@email#1#2{%
 \endgroup
 \patchcmd{\titleblock@produce}
  {\frontmatter@RRAPformat}
  {\frontmatter@RRAPformat{\produce@RRAP{*#1\href{mailto:#2}{#2}}}\frontmatter@RRAPformat}
  {}{}
}%
\makeatother

\begin{document}
\title{Distinguishing Ion Dynamics from Muon Diffusion in Muon Spin Relaxation}
\author{Takashi U. Ito}\thanks{ito.takashi15@jaea.go.jp}
\affiliation{Advanced Science Research Center, Japan Atomic Energy Agency, Tokai-mura, Naka-gun, Ibaraki 319-1195, Japan}
\author{Ryosuke Kadono}\thanks{ryosuke.kadono@kek.jp}
\affiliation{Muon Science Laboratory, Institute of Materials Structure Science, High Energy Accelerator Research Organization (KEK), Tsukuba, Ibaraki 305-0801, Japan}

\begin{abstract}%
We propose a model to describe the fluctuations in the internal magnetic field due to ion dynamics observed in muon spin relaxation ($\mu$SR) by an Edwards--Anderson-type autocorrelation function that separates the quasi-static and dynamic components of the correlation by the parameter $Q$ (where $0\le Q\le1$). 
Our Monte Carlo simulations for this model showed that the time evolution of muon spin polarization deviates significantly from the Kubo--Toyabe function.  To further validate the model, the results of simulations were compared with the $\mu$SR spectra observed in a hybrid organic--inorganic perovskite FAPbI$_3$ [with FA referring to HC(NH$_2)_2$], where local field fluctuations associated with the rotational motion of FA molecules and quasi-static fields from the PbI$_3$ lattice are presumed to coexist. The least-squares curve fitting showed reasonable agreement with the model with $Q=0.947(3)$, and the fluctuation frequency of the dynamical component was obtained. This result opens the door to the possibility of experimentally distinguishing fluctuations due to the  dynamics of ions around muons from those due to the self-diffusion of muons. On the other hand, it suggests the need to carefully consider the spin relaxation function when applying $\mu$SR to the issue of ion dynamics.
\end{abstract}

\maketitle

\section{Introduction}
Muon spin rotation ($\mu$SR) is an experimental method to probe magnetic fields in matter, where the spin-polarized muons ($\mu^+$) stopped in the sample can directly provide information on the magnitude of the local field at the interstitial site(s) via the frequency of their Larmor precession \cite{MSR}.  The muon gyromagnetic ratio $\gamma_\mu$ ($=2\pi\times 135.539$ MHz/T)  is 3.18 times higher than that of protons ($=2\pi\times 42.577$ MHz/T, which is the highest among all the nuclei of stable elements); and thus, it is the most sensitive probe of the internal magnetic field upon their implantation to materials.  Muons are provided as a 100\%-spin-polarized beam by proton accelerator facilities, which enables $\mu$SR measurements in zero magnetic field. In addition, $\mu$SR has various advantages: the muon implantation energy is high enough ($\ge4$~MeV, corresponding to the stopping range $\ge0.1$ g/cm$^2$) to be surface-independent (bulk-sensitive), and the implanted muons (volume concentration $\simeq10^5$ cm$^{-3}$ for the high-flux beams at J-PARC MLF) decay to positrons and neutrinos with an average lifetime of 2.198~$\mu$s; therefore, they do not accumulate in the sample unlike other ion beam irradiation methods.  Most notably, the spatial distribution of the emitted high-energy decay positrons has a large asymmetry (1/3 when averaged over the positron energy) with respect to the muon spin polarization, and the time evolution of spin polarization can be observed by measuring the time-dependent asymmetry (which is called the ``$\mu$SR time spectrum'').

Recently, attempts have been made to observe the diffusive motion of ions in battery materials and ionic conductors by $\mu$SR measurements under zero or longitudinal magnetic field (ZF/LF-$\mu$SR) \cite{Sugiyama:09,Baker:11,Sugiyama:11,Sugiyama:12,Ashton:14, Mansson:14,Amores:16,Laveda:18,Benedek:20,McClelland:20}, where the time-dependent fluctuations of a weak random local field ${\bm H}(t)$ ($\sim$10$^{-4}$ T) due to the magnetic dipole moments of cation nuclei are monitored via the spin relaxation of muons implanted into the materials of interest. In such studies, it is often an issue whether the observed fluctuations are caused by cation diffusion or muon self-diffusion, but it was believed that the distinction could not be made solely from  $\mu$SR spectra. Therefore, the attribution of the origin of the fluctuations has been based on information from other experimental methods and, more recently, on inferences from {\sl ab initio} density functional theory (DFT) calculations.

\begin{figure}[t]
  \centering
\includegraphics[width=0.88\linewidth]{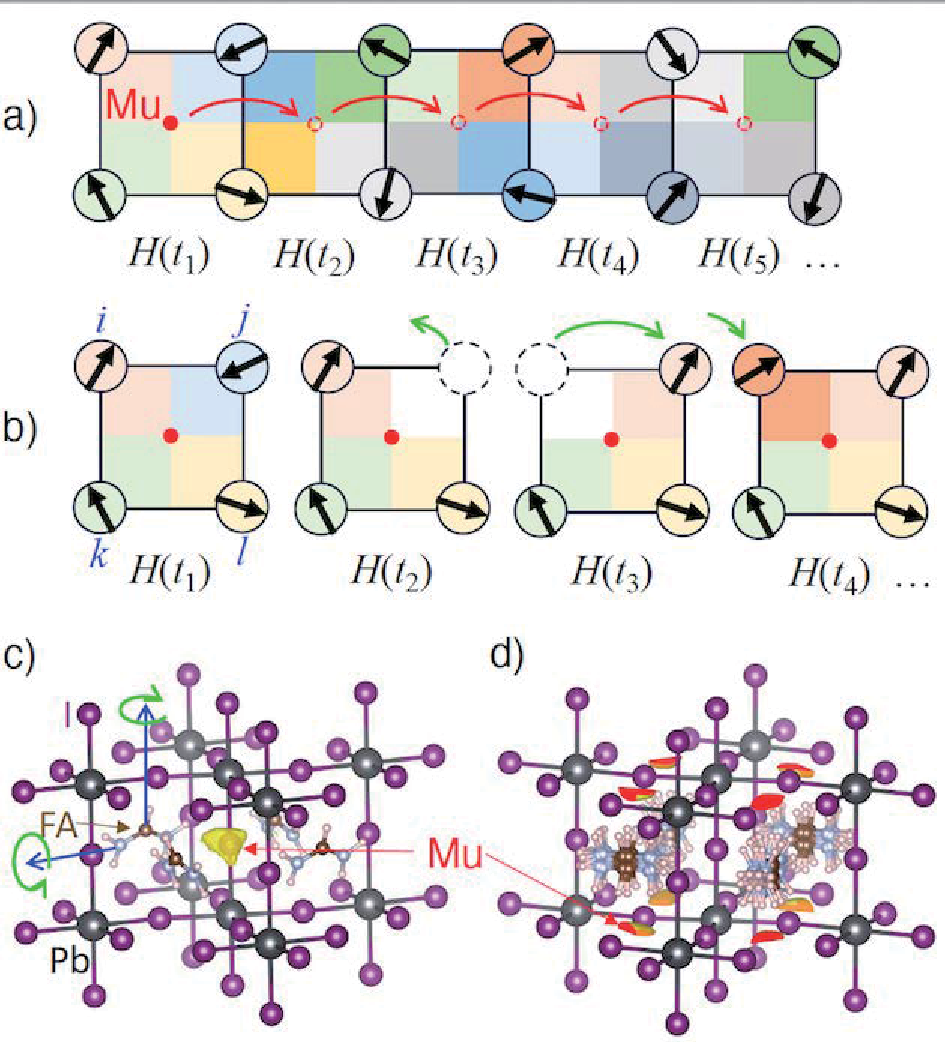}
\caption{(Color online) Schematic of time-dependent fluctuation of local field ${\bm H}(t)$ at muon sites for (a) muon self-diffusion and (b) ion diffusion, where ions are located at corners on a square lattice. The difference in the dipolar field from each ion (with the nuclear magnetic moment orientation shown by an arrow) is shown in different colors (or by brightness in gray scale).
Although the combination of dipolar fields at muon sites changes randomly with muon diffusion in (a), the contributions from two ions at $k$ and $l$ sites remain unchanged, wheras those at $i$ and $j$ sites migrate (b). (c), (d) Structures of formamidinium lead iodate [FAPbI$_3$, with FA denoting HC(NH$_2$)$_2$] in the tetragonal phase, and associated muon sites (areas indicated by arrows) suggested from first-principles DFT calculations (c) for the FA molecules fixed to one of the eight equivalent orientations and (d) for multiple FA orientations using VCA (see Sect.~\ref{comp}). Internal field fluctuations associated with the rotational motion of FA molecules and the static internal field from the PbI$_3$ lattice are presumed to coexist. }
\label{fluc}
\end{figure}

The muon spin relaxation induced by ${\bm H(t)}$ under ZF/LF is considered well approximated by the Kubo--Toyabe (KT) function \cite{Hayano:79}, and it has been routinely used in the analysis of $\mu$SR time spectra by curve fitting based on the least-squares method. In deriving the KT function, the effect of the fluctuations is incorporated using a strong-collision model (an approximation of the Markovian process), where ${\bm H}(t)$ is assumed to have no correlation before and after a change. As illustrated in Fig.~\ref{fluc}(a), this approximation is considered to hold well in the case of the jump diffusion of muons, because the relative configuration of all nuclear magnetic moments is swapped simultaneously in a single jump. On the other hand, in the case of ion diffusion around muons, it is naturally expected to be a rare event that all ions move simultaneously (especially for slow diffusion), resulting in a strong autocorrelation in ${\bm H}(t)$ before and after its change  [see Fig.~\ref{fluc}(b)]. This is even more so when multiple kinds of nuclei contribute to the internal magnetic field (e.g., $^7$Li and $^{59}$Co in Li$_x$CoO$_2$) while only some of them exhibit diffusion. Nevertheless, the KT function has traditionally been used in the study of ion diffusion.

A similar situation has been observed in very recent $\mu$SR studies of hybrid organic--inorganic perovskites (HOIPs) \cite{Koda:22,Hiraishi:23}, where the thermally activated local rotational motion of organic cation molecules contributes to the fluctuation of ${\bm H}(t)$ while the internal fields from the PbI$_3$ lattice remain quasi-static [see Fig.~\ref{fluc}(c)].  Moreover, it has been suggested that the observed $\mu$SR time spectra may not always be satisfactorily reproduced by the KT function upon the onset of molecular motion. These circumstances led us to examine the corresponding longitudinal spin relaxation function for such cases. 

In this paper, we propose a model in which the fluctuations of ${\bm H}(t)$ are described by an Edwards--Anderson-type autocorrelation function with the parameter $Q$ used to divide the quasi-static and dynamical components in the correlation.  Our Monte Carlo simulations on the time evolution of muon spin polarization for this model indicate that the relaxation lineshape could be significantly different from that of the KT function. This difference will open the door to the possibility of  experimentally distinguishing between muon self-diffusion and cation dynamics based solely on $\mu$SR measurements. We show that the origin of ${\bm H}(t)$ fluctuations can indeed be inferred by comparing relaxation functions based on this model with high-precision $\mu$SR time spectra observed in formamidinium lead iodide [FAPbI$_3$, where FA denotes HC(NH$_2)_2$] \cite{Hiraishi:23}. Furthermore, on the basis of this result, we will discuss the problems encountered in the preceding studies of ion diffusion by $\mu$SR.

\section{Model}
\subsection{Relaxation function for the static internal field}
Since our goal is to investigate how different autocorrelations of ${\bm H}(t)$ can change in the spin relaxation function, we will introduce the KT function in a simple classical way in the following.
When muons are initially polarized along the $z$-direction and subjected to a homogeneous magnetic field ${\bm H}=(H_x,H_y,H_z)$, the time evolution of muon polarization is described by
\begin{eqnarray}
\sigma_{z}(t) 
              &=&\frac{H_{z}^{2}}{H^{2}} + 
              \frac{H_{x}^{2}+H_{y}^{2}}{H^{2}}\cos(\gamma_{\mu}Ht)\\
              &=&\cos^2\theta+\sin^2\theta\cos(\gamma_{\mu}Ht),\label{Sz}
\end{eqnarray}
where $H=|{\bm H}|$.
Such a single-frequency rotation is observed when muons are subjected to a uniform external magnetic field (not parallel to $z$). In the case of random local fields exerted from nuclear magnetic moments, $H$ varies with the muon position ${\bm r}_\mu$  so that the $\mu$SR spectrum corresponds to a random sampling of  ${\bm H}({\bm r}_\mu)=(H_x({\bm r}_\mu),H_y({\bm r}_\mu),H_z({\bm r}_\mu))$, which is described by a density distribution:
\begin{equation}
{\bm n}({\bm H})=\int\delta({\bm H}- {\bm H}({\bm r}_\mu))d{\bm r}_\mu.\label{nb}
\end{equation}
The time evolution of spin polarization projected to the $z$-axis is then given by 
\begin{equation}
g_z(t) =  \int\sigma_{z}(t){\bm n}({\bm H})d{\bm H}=\iiint_{-\infty}^{\infty}\sigma_{z}(t)\Pi_\alpha n_\alpha(H_{\alpha})dH_\alpha,\label{Szt}
\end{equation}
where $\alpha=x$, $y$, and $z$.
Unless ${\bm n}({\bm H})$ is a delta function, $g_z(t)$ accompanies relaxation due to the loss of phase coherence in the spin precession.

Specifically, provided that the static distribution of ${\bm H}({\bm r}_\mu)$ is approximated by an isotropic Gaussian distribution,
\begin{equation}
n_\alpha(H)=\frac{\gamma_{\mu}}{\sqrt{2\pi}\Delta}
\exp\left(-\frac{\gamma_{\mu}^{2}H^{2}}{2\Delta^{2}}\right),  \:\:   (\alpha=x,y,z),
\label{ph}
\end{equation}
the corresponding relaxation function is obtained analytically by executing the integral in Eq.~(\ref{Szt}) to yield the  {\sl static} KT function \cite{Hayano:79},
\begin{equation}
g_{z}(t) =   \frac{1}{3}+\frac{2}{3}(1-\Delta^{2}t^{2})e^{-\frac{1}{2}\Delta^{2}t^{2}}, \label{gkt}
\end{equation}
where the term $1/3$ originates from the spatial average of $\cos^2\theta$ in Eq.~(\ref{Sz}), corresponding to the probability that ${\bm H}({\bm r}_\mu)$ is parallel to $z$.
The linewidth $\Delta$ is determined by the second moment of the dipole field from the nuclear magnetic moments of the neighboring atoms (including noncationic ions),
\begin{equation}
\Delta^2=\frac{1}{2}\gamma_\mu^2\gamma_I^2\sum_i\sum_{\alpha=x,y}\sum_{\beta=x,y,z}(\hat{A}_i^{\alpha\beta}{\bm I}_i)^2, \label{delta_n}
\end{equation}
where 
\begin{equation} 
(\hat{A}_i)^{\alpha\beta}
=\frac{1}{r^3_i}(\frac{3\alpha_i \beta_i}{r^2_i}-\delta_{\alpha\beta}) \:\:(\alpha,\beta=x,y,z)\label{diptensor}
\end{equation}
is the dipole tensor for the $i$th nuclear magnetic moment $\gamma_I{\bm I}_i$ situated at ${\bm r}_i$ from the muon site. Since the magnetic dipolar field decays proportionally to $1/r_i^3$, the magnitude of $\Delta$ is approximately determined by the nearest neighboring (nn) nuclear magnetic moments around a muon. The Gaussian distribution for ${\bm n}({\bm H})$ works reasonably well when the number $N$ of nn nuclear magnetic moments satisfies the condition $N\ge4$. On the other hand, it is a relatively poor approximation for $N\le3$, and it is necessary to calculate $g_z(t)$ by treating the muon--nucleus cluster as a few-quantum spin system using the density matrix method \cite{Brewer:86,Celio:86}. With recent advances in the computational environment, such calculations have been performed for nuclear magnetic moments located farther from a muon, and good agreement of calculation results with experimental results has been obtained \cite{Wilkinson:20}.

\subsection{Effect of fluctuating internal fields}

In general, the effect of fluctuations in linear response theory is treated by a Gaussian--Markov process. In this model, the autocorrelation of fluctuations is given by 
\begin{equation}
\frac{\langle {\bm H}(t){\bm H}(0)\rangle}{\langle H(0)^2\rangle}= \frac{\langle {\bm H}(t){\bm H}(0)\rangle}{\Delta^2/\gamma_\mu^2}=e^{-\nu t}, \label{Acf}
\end{equation}
where $\langle...\rangle$ denotes the thermal average over the canonical ensemble. (For a more rigorous treatment, see, for example, Ref.~\onlinecite{Takahashi:20}.)  The strong-collision model (also called the random Markov process or random phase approximation) is a simplified version of this model, where it is assumed that the density distribution $n_\alpha(H)$ is constant at any time $t$ and that ${\bm H}(t)$ changes with an average rate $\nu=1/\tau$ without correlation before and after a change in ${\bm H}$.  The resulting relaxation function is a good approximation of a Gaussian process except in the $\nu t\ll1$ region \cite{Kehr:78}.
In this case, the relaxation function $G_z(t)$, which incorporates the effect of fluctuations on the static relaxation function $g_z(t)$, is derived by solving the following integral equation (a linear Volterra equation of the second kind):
\begin{equation}
G_z(t)=e^{-\nu t}g_z(t)+\nu\int_0^t d\tau e^{-\nu(t-\tau)}g_z(t-\tau)G_z(\tau).\label{Gdyn}
\end{equation}
This integral equation can be solved analytically by the Laplace transform of $g_z(t)$ and $G_z(t)$  \cite{Hayano:79} or direct numerical integration \cite{Reotier:92}.
The dynamical KT function $G^{\rm KT}_z(t;\Delta,\nu)$ obtained in this way successfully reproduced the spin relaxation due to muon jump diffusion in a nonmagnetic metal \cite{Clawson:83,Kadono:89,Luke:91}. This indicates that fluctuations in the internal magnetic field due to muon self-diffusion can be well described by the strong-collision model with Eq.~(\ref{Acf}).

On the other hand, fluctuations in ${\bm H}$ due to ion diffusion do not necessarily occur in its entire amplitude; let us consider the situation in Fig.~\ref{fluc}(b) where part of ${\bm H}$ is due to contributions from cations or anions that remain stationary while other ions exhibit local motion. 
A similar situation is expected in the case of FAPbI$_3$ shown in Fig.~\ref{fluc}(c), where cation molecules are in random rotational motion relative to the static PbI$_3$ lattice.
In these cases, it is clear that the autocorrelation function differs from Eq.~(\ref{Acf}), and it may be given by
\begin{equation}
\frac{\langle {\bm H}(0){\bm H}(t)\rangle}{\Delta^2/\gamma_\mu^2}\approx (1-Q)e^{-\nu_\mu t}+Qe^{-(\nu_{\rm i}+\nu_\mu) t}, \label{AcfEA}
\end{equation}
where $Q$ is the parameter that describes the fractional amplitude exhibiting fluctuations with the rate $\nu_{\rm i}$, and 
$\nu_\mu$ is the fluctuation rate due to muon diffusion.  
The equation with $\nu_{\rm \mu}=0$ corresponds to a model introduced to describe the slowing down of magnetic fluctuations (with the rate $\nu_{\rm i}$) in spin glasses, where $1-Q$ is interpreted as an order parameter that describes ``ordering in time'' \cite{Edwards:75,Edwards:76}. This model was also applied to deriving relaxation functions for analyzing the $\mu$SR spectra observed in dilute-alloy spin glasses \cite{Uemura:85}. 
\begin{figure}[t]
  \centering
\includegraphics[width=0.95\linewidth]{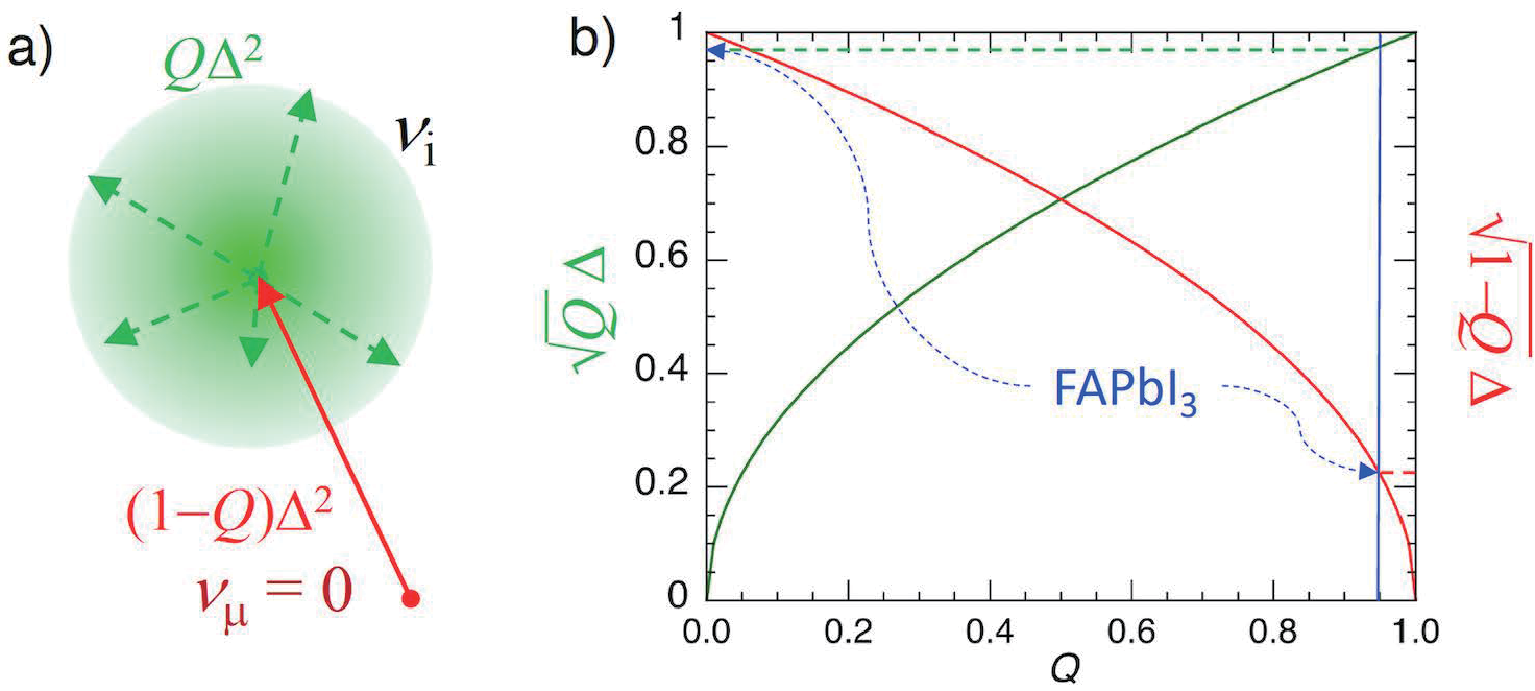}
\caption{(Color online) (a) Schematic of local fields consisting of static and fluctuating components, where $Q$ is the parameter representing the fractional amplitude of the local fields  fluctuating at a frequency $\nu_i$. (b) $Q$ dependence of the linewidths of the dynamical ($\sqrt{Q}\Delta$) and static ($\sqrt{1-Q}\Delta$) components corresponding to (a) [$\Delta=1$, see Eq.~(\ref{AcfEA2}) in the main text]. Dotted arrows correspond to the case for FAPbI$_3$ ($Q=0.947$).}
\label{EA}
\end{figure}

However, the corresponding $G_z(t)$ has only been discussed in approximate forms obtained as analytical expressions when $\nu_\mu\ll\Delta\ll\nu_{\rm i}$ \cite{Koda:22,Uemura:85}, and an extension to the general case is still awaited. Therefore, in the following, we investigate the behavior of $G_z(t)$ corresponding to arbitrary $Q$ and $\nu_i$  by Monte Carlo simulations.

\subsection{Monte Carlo simulations}
We performed numerical simulations of the spin relaxation function for $Q>0$ and $\nu_\mu=0$, which we call $G^{\rm ID}_z(t;\Delta,\nu_{\rm i},Q)$, where the ion dynamics is the sole origin of fluctuations with the amplitude $Q\Delta^2$ and the frequency $\nu_{\rm i}$: 
\begin{equation}
 \gamma_\mu^2\langle{\bm H}(0){\bm H}(t)\rangle = [\sqrt{1-Q}\Delta]^2 + [\sqrt{Q}\Delta]^2e^{-\nu_{\rm i} t}.\label{AcfEA2}
\end{equation}
Figure \ref{EA} schematically shows the situation in Eq.~(\ref{AcfEA2}) under these conditions. In our Monte Carlo simulation, the local field that each muon experiences was approximately expressed as a vector sum of static and dynamical fields. These fields were randomly generated according to the Gaussian probability distribution $n_\alpha(H)$ with the widths of $\sqrt{(1-Q)}\Delta/\gamma_\mu$ and $\sqrt{Q}\Delta/\gamma_\mu$. 
Note that the assignment of $1-Q$ and $Q$ is opposite to that in Refs.~\onlinecite{Edwards:75} and \onlinecite{Edwards:76}, where $Q$ represents the {\sl static} component; this is so defined such that the case $Q=0$ should be reduced to the conventional KT function (with $\nu_\mu>0$ for muon diffusion). The fluctuation of the dynamical component was implemented in accordance with the strong-collision model by regenerating the dynamical field at the average rate of $\nu_{\rm i}$. The time evolution of a muon spin, initially oriented to the $z$-direction, in the local field was calculated and its $z$-component was averaged for $10^8$ muons to obtain $G_z^{\rm ID}(t)$ for ZF. For finite LFs, the  time evolution was calculated in the field obtained by the vector sum of the local field and the LF along the $z$-direction.

\begin{figure}[t]
\centering
\includegraphics[width=0.9\linewidth]{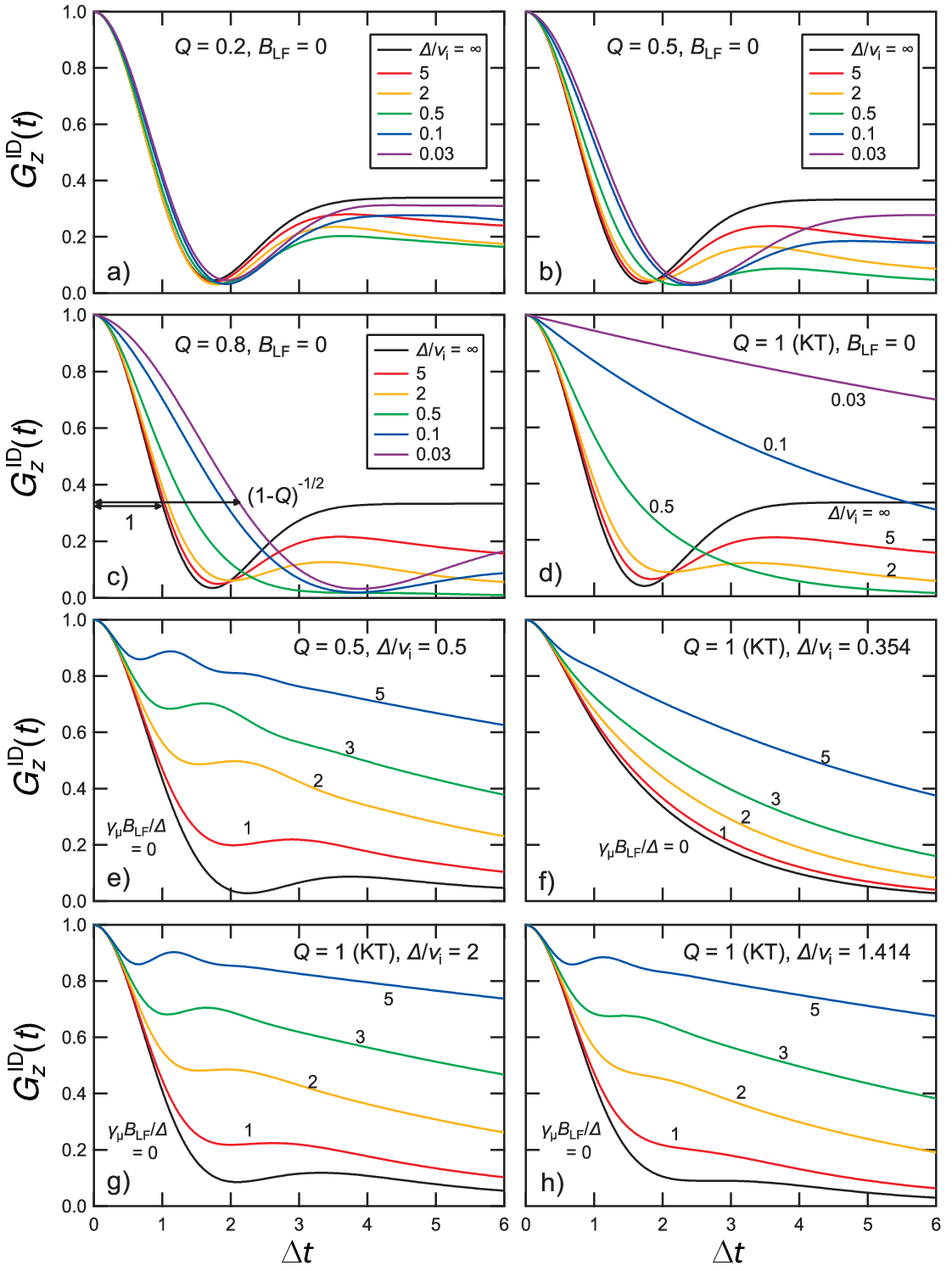}
\caption{(Color online) Some examples of the muon spin relaxation function $G^{\rm ID}_z(t;\Delta,\nu_{\rm i},Q)$ incorporating the effect of ion dynamics simulated using the parameter $Q$, where $Q\Delta^2$ is the amplitude of fluctuation. (a)--(d) under zero field with different $Q$ and $\nu_{\rm i}$ values, and (e)--(h) under various longitudinal fields $B_{\rm LF}$ with $Q=0.5$ and $\Delta/\nu_{\rm i}=0.5$. Note that $Q=1$ (and 0) corresponds to the conventional KT function. }
\label{Gzsim}
\end{figure}

As shown in Figs.~\ref{Gzsim}(a)--\ref{Gzsim}(c), in the case of $0<Q<1$, the $\frac{1}{3}$ term once decays as $\nu_{\rm i}$ increases, then begins to recover with the apparent linewidth decreasing to $\sqrt{1-Q}\Delta$. Provided that the $\mu$SR spectrum showing such temperature dependence is analyzed with the usual KT function [Fig.~\ref{Gzsim}(d)], the apparent fluctuation rate would increase and then decrease. Thus, it is likely to lead to an incorrect interpretation that the cause of the fluctuations is different from a simple thermal excitation process. An important clue to avoiding this is to determine whether the linewidth changes at both ends of the apparent increase and decrease in $\nu$ derived from the KT function. In Figs.~\ref{Gzsim}(b) and \ref{Gzsim}(c), it can be seen that as $\nu_{\rm i}$ increases, the $Q\Delta^2$ contribution disappears owing to motional narrowing, and the relaxation rate decreases (converging to $\sqrt{1-Q}\Delta$) while maintaining the Gaussian linshape. Such behavior is actually observed in HOIP \cite{Koda:22,Hiraishi:23}, and the change in the overall lineshape is found to correspond approximately to the case of $Q\approx0.95$ (see below). Note here that the linewidth behaves nonlinearly with respect to $Q$. As shown in Fig.~\ref{EA}(b), it changes steeply around $Q\rightarrow0$ for $\sqrt{Q}\Delta$ and $Q\rightarrow1$ for $\sqrt{1-Q}\Delta$.

The LF dependence of $G^{\rm ID}_z(t;\Delta,\nu_{\rm i},Q)$ is indispensable information for a more accurate determination of fluctuation frequency in $\mu$SR measurements. Let us compare the field dependence of the time spectrum for $Q=0.5$ and $\Delta/\nu_{\rm i}=0.5$ shown in Fig.~\ref{Gzsim}(e) with that of the conventional KT function in detail. First, the KT function calculated for $\Delta/\nu_{\rm i}=\frac{1}{2}\sqrt{0.5}$ is shown in Fig.~\ref{Gzsim}(f); this is because the linewidth for the fluctuations in the former case is $\sqrt{Q}\Delta$ rather than $\Delta$. The oscillations around $\Delta\cdot t\simeq2$ [originating from the static component corresponding to $(1-Q)$] seen in Fig.~\ref{Gzsim}(e) are absent, indicating a qualitatively different behavior. Furthermore, assuming practical situations of curve fitting with the KT function, the results with freely varying $\nu_{\rm i}$ to reproduce the behavior in Fig.~\ref{Gzsim}(e) are shown in Figs.~\ref{Gzsim}(g) and \ref{Gzsim}(h). Figure \ref{Gzsim}(g) reproduces the low-field spectrum well, whereas the fit is poor on the high-field side; Fig.~\ref{Gzsim}(h) shows the opposite trend. From these comparisons, we can conclude that the curve fitting with the KT function does not accurately reproduce the field dependence of the spectrum when $Q<1$.

Now, let us discuss the behavior of the relaxation function for some extreme cases, for which we define $\Delta_\mu=\sqrt{(1-Q)}\Delta$ and $\Delta_{\rm i}=\sqrt{Q}\Delta$. When the ion dynamics is fast enough to satisfy $\nu_{\rm i}\gg\Delta$, the spin relaxation for that part is approximated to
\begin{equation}
G_{\rm i}(t)\simeq G_z^{\rm KT}(t;\Delta_{\rm i},\nu_{\rm i})=e^{-\frac{2\Delta_{\rm i}^2t}{\nu_{\rm i}}}.
\end{equation}
Then, if the muon is quasi-static, the corresponding total relaxation function is
\begin{equation}
G^{\rm ID}_z(t)\simeq G_z^{\rm KT}(t;\Delta_\mu,0)G_{\rm i}(t)=[\frac{1}{3}+\frac{2}{3}(1-\Delta_\mu^{2}t^{2})e^{-\frac{1}{2}\Delta_\mu^{2}t^{2}}]e^{-\frac{2\Delta_{\rm i}^2t}{\nu_{\rm i}}}, 
\end{equation}
indicating that the curve fitting using the conventional KT function will yield a reduced linewidth $\sqrt{(1-Q)}\Delta$. 

Conversely, in the limit of slow ion dynamics ($\nu_{\rm i}\ll\Delta$), the change in relaxation function is independent of linewidth, and the exponential relaxation of the $\frac{1}{3}$ component dominates:
\begin{equation}
G^{\rm ID}_z(t)\simeq \frac{1}{3}e^{-\nu_{\rm i}t}+\frac{2}{3}(1-\Delta^{2}t^{2})e^{-\frac{1}{2}\Delta^{2}t^{2}},
\end{equation}
which is valid for $Q\approx1$. This suggests that it is difficult to distinguish the ion dynamics from muon diffusion in the slow exponential damping. As demonstrated in the case of FAPbI$_3$ below, the difference from the conventional KT function becomes notable in the intermediate region of $\nu_{\rm i}\gtrsim\Delta$, so the spectra in such a temperature region are important in data analysis to distinguish between ion dynamics and muon self-diffusion. 

\subsection{Comparison with experimental results}\label{comp}
One of the fields of $\mu$SR application that would greatly benefit from the possible distinction between local ion dynamics and muon diffusion would be in the study of ion diffusion. From the viewpoint of the present model, however, we are still unable to find convincing examples of $\mu$SR data that exhibit the characteristic behavior of ion diffusion revealed in our simulations. 

For instance, an earlier $\mu$SR study on
Li$_{0.73}$CoO$_2$ reported that Li ions exhibit diffusive motion whereas $\mu^+$ remains bound to an O atom and immobile at low temperatures \cite{Sugiyama:09}. Above $T\approx300$ K,  the linewidth obtained by KT function analysis ($\equiv\Delta_{\rm KT}$) shows a significant decrease with increasing $T$, which is attributed to the onset of muon diffusion in the literature. Since such a narrowing behavior is also expected to occur in the limit of large $\nu_{\rm i}$ ($\Delta/\nu_{\rm i}\rightarrow0$) in our model [see Figs.~\ref{Gzsim}(a)--\ref{Gzsim}(c)], let us examine the consistency of our model with the reported $\mu$SR result across the entire $T$ range from 10 to 400 K.

\begin{figure}[t]
\centering
\includegraphics[width=0.78\linewidth]{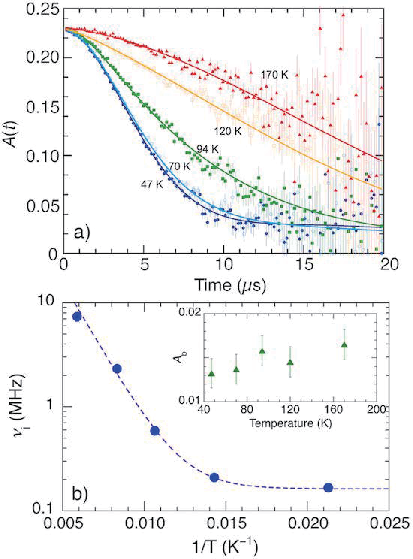}
\caption{(Color online) (a) Zero-field $\mu$SR spectra observed at low temperatures in FAPbI$_3$ fitted with a relaxation function calculated by Monte Carlo simulation (global fitting assuming that the initial asymmetry, $Q$, and $\Delta$ are common for all temperatures). All the spectra are reproduced by using $Q=0.947(3)$ and $\Delta=0.191(1)$ MHz. (b) Arrhenius plot of the internal field fluctuation frequency $\nu_{\rm i}$ attributed to the FA molecular motion obtained in this analysis. The dashed line is the fitting by the sum of the thermally activated function and the constant. Inset: temperature dependence of the constant background (see text for details).
}
\label{fapbi}
\end{figure}

From the value of $\Delta_{\rm KT}$ extrapolated to $T\rightarrow0$~K and that to $\sim$400 K \cite{Sugiyama:09}, the total linewidth $\Delta$ and its static part $\sqrt{1-Q}\Delta$ in the limit of $\nu_{\rm i}\rightarrow\infty$ for our model are respectively estimated to be 0.26 and 0.07 MHz, yielding $Q=1-(0.07/0.26)^2=0.93$. However, this value disagrees with that obtained by another evaluation, $Q=1-(0.23/0.43)^2=0.71$ derived from $\Delta_{\rm KT}=0.23$ and 0.43 MHz respectively calculated for $x=0$ and 0.75 in Li$_x$CoO$_2$ at the presumed muon site adjacent to oxygen \cite{Sugiyama:09,Mansson:13,Sugiyama:15}. Moreover, in our model with $Q=0.93$, ZF-$\mu$SR time spectra should retain Gaussian-like features for any $\Delta / \nu_{\rm i}$ value, as exemplified in 
Fig.~\ref{fapbi}. On the other hand, the spectrum observed at 275 K \cite{Mansson:13,Sugiyama:15} exhibits a exponential lineshape similar to that for $\Delta/\nu_{\rm i}<0.5$ in Fig.~\ref{Gzsim}(d) ($Q=1$, where $\Delta=\Delta_{\rm KT}$), and the lineshape is suggested to become more exponential above 275 K as inferred from the decrease in $\Delta/\nu_{\rm i}$ to $\sim$0.08 at 400 K. Thus, it is difficult to explain the behavior of the ZF-$\mu$SR spectra in Li$_{0.73}$CoO$_2$ over the entire $T$ range in a unified manner with our immobile-$\mu$ model, and it may be necessary to consider the effect of muon diffusion at lower temperatures, as has been discussed recently \cite{Ohishi:22}.

On the low $Q$ side, our simulation indicates that the change in lineshape is strongly suppressed; as seen in Fig.~\ref{Gzsim}(a) ($Q=0.2$), the time spectral change is so small that it appears to correspond to quasi-static KT functions regardless of $\Delta/\nu_{\rm i}$. This still seems inconsistent with the significant time spectral change  (and associated $\nu$) in Li$_x$CoO$_2$. Besides this, such small deviations from the KT function for a small $Q$ are also expected from other causes, e.g., a smaller number of nearest-neighbor nuclear magnetic moments around a muon, making it more difficult to identify the true cause of the spectral change. In contrast, as discussed below,  FAPbI$_3$ turns out to be an excellent example for testing our model because it suggests a large $Q$.

It has been inferred from a previous $\mu$SR study that muons in FAPbI$_3$ are exposed to random local fields originating from the nuclear magnetic moments of both FA molecules and the PbI$_3$ lattice \cite{Hiraishi:23}. Among these fields, the magnetic field from the FA molecule (mainly due to the proton magnetic moment) is expected to exhibit fluctuations due to the rotational motion induced by thermal excitation. The analysis of the $\mu$SR time spectra in the tetragonal phase of FAPbI$_3$ using the KT function showed that the linewidth $\Delta$ appears to decrease monotonically from $\sim$50 to 140 K, whereas the fluctuation frequency $\nu$ exhibits a weak peak at around 100 K. Moreover, the global curve fitting including the spectra under various LFs are relatively poor in the 50--120 K range where $\Delta$ changes significantly.  These observations point to the situation that the actual relaxation function deviates from the KT function, as demonstrated in Fig.~\ref{Gzsim}. Therefore, we focused on the temperature variation of the ZF-$\mu$SR spectrum and investigated whether the spectral changes could be reproduced by using the appropriate $Q$, $\Delta$, and $\nu_{\rm i}$ when the data were analyzed with the new model.  

Specifically, we prepared a multidimensional numerical table of $G^{\rm ID}_z(t;\Delta,\nu_{\rm i},Q)$ for the parameters $\Delta\cdot t$, $Q$, and $\nu_{\rm i}$ calculated by Monte Carlo simulation \cite{SM}, and we performed the least-squares curve fitting of the time-dependent asymmetry using the equation
\begin{equation}
A(t)=A_0G^{\rm ID}_z(t;\Delta,\nu_{\rm i},Q)+A_{\rm b},
\end{equation}
where $A_0$ is the initial asymmetry depending on the instrument and $A_{\rm b}$ is the background signal from muons that missed the sample. In addition, a global fitting for multiple spectra was performed, where $A_0$, $\Delta$, and $Q$ were common for the spectra at all temperatures. As shown in Fig.~\ref{fapbi}(a), all the spectra were well reproduced (reduced net chi-square = 1.33), and the values of the two common parameters were obtained as $Q=0.947(3)$ and $\Delta=0.191(1)$ MHz.  

The theoretical evaluation of $\Delta$ in the tetragonal phase of FAPbI$_3$ has a problem in that it is necessary to take into account the multiple orientations of FA molecules in the PbI$_3$ lattice, which makes the {\sl ab initio} DFT calculations difficult in the Tetra-LT phase ($<140$ K). Therefore, we discuss the results of DFT calculations for the Tet-HT phase (140--280 K), where the contribution of FA molecules to $\Delta$ is quenched by the motional effect. More specifically, the hydrogen defect calculations (for mimicking muons) were conducted in two cases, one with the FA molecules fixed to one of the eight equivalent orientations and the other with all possible orientations taken into account using the virtual crystal approximation (VCA). (See Ref.~\onlinecite{Hiraishi:23} for the details of the DFT calculations.) The predicted muon sites in each case are shown in Figs.~\ref{fluc}(c) and \ref{fluc}(d), and the corresponding $\Delta$ values are summarized in Table \ref{Dlt}. Although the agreement between the calculation and experimental results is not necessarily good for these absolute values, the result of VCA calculation is in reasonable agreement with experimental result for $\sqrt{1-Q}$, which indicates the relative change in $\Delta$ with and without the contribution of FA molecules.

\begin{table}
\begin{center}
\begin{tabular}{c|cc|c}
\hline\hline
 & \hspace{1em}Fixed FA\hspace{1em}  & \hspace{1em}VCA\hspace{1em} & \multirow{2}{*}{\hspace{2em}Exp.\hspace{2em}}\\
 Muon site & $16l$  & $2d$ & \\
 \hline
 $\Delta$ (MHz) & 0.179 &0.2941 & 0.191(1)\\
 $\Delta_{\rm PbI_3}$ (MHz) & 0.1022 & 0.0805 & {\it 0.044(1)$^*$}\\
\hline
 $\sqrt{1-Q}$ ($=\Delta_{\rm PbI_3}/\Delta$) & 0.571 & 0.274 & 0.230(1)\\
\hline\hline
\multicolumn{4}{l}{$^*$calculated from $\Delta=0.191$ MHz and $\sqrt{1-Q}=0.230$.}  
\end{tabular}
\caption{Nuclear dipolar linewidth calculated for the muon sites inferred from DFT calculations, where $\Delta$ is the total linewidth and $\Delta_{\rm PbI_3}$ is that without the contribution of FA molecules \cite{Hiraishi:23}. ``Fixed FA'' and ``VCA'' correspond to those shown in Figs.~\ref{fluc}(c) and \ref{fluc}(d), respectively, where the FA molecular orientation was set to be common in the calculations of $\Delta$. }\label{Dlt}
\end{center}
\end{table}

More importantly, as shown in Fig.~\ref{fapbi}(b), we successfully demonstrated using $\mu$SR that $\nu_{\rm i}$ exhibits a temperature dependence consistent with the thermally activated process in the temperature range where the rotational motion of FA molecules is strongly suggested from other experiments; this was not obtained directly by analysis using the KT function \cite{Hiraishi:23}. Curve fitting using the Arrhenius function, $\nu_{\rm i}=\nu_0\exp(-E_{\rm a}/k_BT) + \nu_c$, yields excellent agreement with $\nu_0=3.6(6)\times10^8$ s$^{-1}$, $E_{\rm a}/k_B=630(18)$ K, and $\nu_c=1.62(9)\times10^5$ s$^{-1}$ [see the dashed line in Fig.~\ref{fapbi}(b)].  $A_{\rm b}$ is small ($\simeq0.015\ll A_0$) and nearly independent of temperature as expected, supporting the present model. 

\section{Discussion}
We have shown by Monte Carlo simulations that when the fluctuations of ${\bm H}(t)$ due to ion dynamics correspond to $0<Q<1$ in Eq.~(\ref{AcfEA2}), we encounter a $\Delta/\nu_{\rm i}$ region where analysis using the KT function provides unsatisfactory results. This immediately points to the possibility that a careful analysis of the time spectra in the temperature region showing such $\Delta/\nu_{\rm i}$ may enable us to determine the origin of the fluctuations.  We then compared our simulation data with previous $\mu$SR experimental data for the Li ion battery material Li$_x$CoO$_2$  and the hybrid perovskite FAPbI$_3$. As a result, it was demonstrated in the latter that the lineshapes of the time spectra are coherently reproduced by the simulations and that a reasonable temperature dependence of the cation molecular fluctuation frequency is obtained, whereas it was suggested in the former that the experimental results exhibit a qualitatively different behavior from the simulations.

Historically, the possibility of applying $\mu$SR to the study of ion dynamics was first discussed for the diffusion of Li$^+$ ions in Li$_x$CoO$_2$ \cite{Sugiyama:09}, and then the application of $\mu$SR to other transition metal oxides containing Li has been attempted \cite{Baker:11,Sugiyama:11,Sugiyama:12,Ashton:14, Mansson:14,Amores:16,Laveda:18,Benedek:20,McClelland:20}. In such studies, the $\mu$SR time spectra reflecting ${\bm H(t)}$ originating from the $^7$Li and other nuclear magnetic moments (e.g., $^{59}$Co in  Li$_x$CoO$_2$) were analyzed using the dynamical KT function, and it was commonly assumed that the obtained fluctuation rate $\nu$ directly corresponds to the jumping frequency of Li$^+$ ions. 

However, a very recent paper reporting the results of an ``operando'' $\mu$SR experiment in the charging and discharging of Li$_x$CoO$_2$ incorporated in a half battery suggests an alternative interpretation that the fluctuation rate $\nu$ (derived from the analysis using the KT function) reflects the contributions of both Li$^+$ and muon jump diffusion, and that $\nu$ divided by the square root of the Li--muon mass ratio ($\sqrt{m_{\rm Li}/m_\mu}\simeq7.9$) is the jump frequency of Li \cite{Ohishi:22}.  In this regard, it is worth mentioning the established fact that muons diffuse in crystalline solids via the thermally activated tunneling process at temperatures higher than about a fraction of the Debye temperature $\Theta_D$ \cite{Miyazaki:03,Fukai:05}. Although the above interpretation is based on different muon sites inferred from simplistic DFT calculations \cite{Ohishi:22}, we need to consider the general possibility that muon diffusion via the thermally activated tunneling process prevails over ion diffusion in the relevant temperature range.

Another important factor in terms of identifying the origin of fluctuation is the temperature dependence of $\Delta$ derived from the analysis using the KT function. Although the decrease in apparent $\Delta$ with increasing temperature may indicate a contribution of ion dynamics to spin relaxation, muon diffusion is also expected in the temperature region of interest. Our simulation results show that when the muon diffusion is dominant ($\nu_\mu\gg\nu_{\rm i}$), the spin relaxation can be well reproduced using the KT function. Therefore, when the spectra are well reproduced using the KT function even though $\Delta$ exhibits a change (as seen in the case of Li$_x$CoO$_2$), it is more likely that the cause of the change in $\Delta$ is attributed to the issue of muon site(s) rather than that of ion dynamics; note that $\Delta$ can vary with the muon position ${\bm r}_\mu$ in inhomogeneous materials. For example, muons may prefer to occupy atomic vacancies or voids to reduce zero-point energy, which has a significant impact on the behavior of light particles \cite{Ito:23}.  Provided that $\Delta=\Delta({\bm r}_\mu)$ is small for such defect sites, the mean $\Delta$ will decrease as the probability of a muon reaching these sites increases with temperature in a diffusion-limited process. Such behavior has been actually observed, for example, in InGaZnO \cite{Kojima:19} and VO$_2$ \cite{Okabe:24}.

In examining previously reported $\mu$SR studies of Li ion diffusion, there are many cases in which $\Delta$ decreases with increasing temperature, but thus far there has been little discussion on such behavior from the aforementioned viewpoint. If the observed spin relaxation is well reproduced by the KT function, it is incompatible with the ion diffusion for which the autocorrelation function is assumed to be given by Eq.~(\ref{AcfEA2}) with $Q<1$. To determine whether the observed fluctuations are due to ion diffusion, it will be necessary to examine the reproducibility of $\mu$SR spectra using the model presented here with $Q<1$.

Finally, as a more general possibility, note that $Q$ may not necessarily be constant over the temperature range of interest. For example, there could be a situation where the diffusion motion of two kinds of ions (other than $\mu^+$) occurs in stages as the temperature increases because of different activation energies. Developing a model that takes such a possibility into account is a future issue.

\section{Conclusion}
We have performed Monte Carlo simulations to investigate how the residual correlations in the fluctuations described by the Edwards--Anderson type parameter $Q$ affect the lineshape of the relaxation function observed in the $\mu$SR experiment. As a result, we found a significant deviation from the KT function for $0<Q<1$. We then compared the simulated relaxation function, $G_z^{\rm ID}(t;\Delta,\nu_{\rm i},Q)$, with the $\mu$SR spectra observed in FAPbI$_3$ in which internal field fluctuations associated with the rotational motion of FA molecules and the quasi-static internal field from the PbI$_3$ lattice are presumed to coexist. The curve fitting shows that the model can provide an appropriate description of the temperature-dependent lineshape with reasonable $\Delta$, $\nu_{\rm i}$ and $Q$, which is in line with cation molecular motion. This result paves the way for the possibility of experimentally distinguishing fluctuations due to ion dynamics around muons from muon self-diffusion, which has been considered difficult. It also calls into question the attribution of internal field fluctuations entirely to ion diffusion for the $\mu$SR time spectra well reproduced by the KT function, suggesting the need for a major reexamination of the results of previous analyses.

\section*{Acknowledgment}
We thank M. Hiraishi, H. Okabe, A. Koda, and K. Fukutani for their helpful discussion during the preparation of the manuscript. The $\mu$SR results for FAPbI$_3$ quoted in this paper were those published in Ref.~\onlinecite{Hiraishi:23}. The Monte Carlo simulation was conducted using the supercomputer HPE SGI8600 in Japan Atomic Energy Agency.
This work was partially supported by JSPS KAKENHI (Grant Nos. 23K11707, 21H05102, and 20H01864) and the MEXT Program: Data Creation and Utilization Type Material Research and Development Project (Grant No. JPMXP1122683430).

\let\doi\relax
%

\end{document}